\documentclass[sigconf]{acmart}
\usepackage{multirow}
\usepackage{balance}

\usepackage[T1]{fontenc}
\usepackage{aecompl}
\AtBeginDocument{%
  \providecommand\BibTeX{{%
    \normalfont B\kern-0.5em{\scshape i\kern-0.25em b}\kern-0.8em\TeX}}}

\setcopyright{acmcopyright}
\copyrightyear{2022}
\acmYear{2022}
\setcopyright{acmcopyright}
\acmConference[WSDM '22] {Proceedings of the Fifteenth ACM International Conference on Web Search and Data Mining}{February 21--25, 2022}{Tempe, AZ, USA.}
\acmBooktitle{Proceedings of the Fifteenth ACM International Conference on Web Search and Data Mining (WSDM '22), February 21--25, 2022, Tempe, AZ, USA}
\acmPrice{15.00}
\acmISBN{978-1-4503-9132-0/22/02}
\acmDOI{10.1145/3488560.3498522}

\begin{document}
\fancyhead{}
\title{Community Trend Prediction on Heterogeneous Graph in E-commerce}

\settopmatter{printacmref=false, printfolios=false}
\author{
Jiahao Yuan\textsuperscript{\rm 1}\textsuperscript{*},
Zhao Li\textsuperscript{\rm 2}\textsuperscript{*}, Pengcheng Zou\textsuperscript{\rm 2}, Xuan Gao\textsuperscript{\rm 2}, Jinwei Pan\textsuperscript{\rm 1}, Wendi Ji\textsuperscript{\rm 1}\textsuperscript{\dag},  Xiaoling Wang\textsuperscript{\rm 1}\textsuperscript{\rm 3}}
\thanks{\textsuperscript{*} Equal contribution}
\thanks{\textsuperscript{\dag} Corresponding author}

\affiliation{
\institution{\textsuperscript{\rm 1} Shanghai Key Laboratory of Trustworthy Computing, East China Normal University \country{China}}
}
\affiliation{
\institution{\textsuperscript{\rm 2} Alibaba Group \country{China}}}
\affiliation{
\institution{\textsuperscript{\rm 3} Shanghai Institute of Intelligent Science and Technology, Tongji University \country{China}}}
\email{
  {jhyuan,wjpan}@stu.ecnu.edu.cn, {wdji,xlwang}@cs.ecnu.edu.cn
}
\email{
 {lizhao.lz, xuanwei.zpc, jingyao.gx}@alibaba-inc.com
}

\fancyhead{}
\renewcommand{\shortauthors}{Jiahao Yuan and Zhao Li, et al.}

\begin{abstract}
In online shopping, ever-changing fashion trends make merchants need to prepare more differentiated products to meet the diversified demands, and e-commerce platforms need to capture the market trend with a prophetic vision.
For the trend prediction, the attribute tags, as the essential description of items, can genuinely reflect the decision basis of consumers. 
However, few existing works explore the attribute trend in the specific community for e-commerce. 
In this paper, we focus on the community trend prediction on the item attribute and propose a unified framework that combines the dynamic evolution of two graph patterns to predict the attribute trend in a specific community.
Specifically, we first design a community-attribute bipartite graph at each time step to learn the collaboration of different communities. 
Next, we transform the bipartite graph into a hypergraph to exploit the associations of different attribute tags in one community. 
Lastly, we introduce a dynamic evolution component based on the recurrent neural networks to capture the fashion trend of attribute tags. 
Extensive experiments on three real-world datasets in a large e-commerce platform show the superiority of the proposed approach over several strong alternatives and demonstrate the ability to discover the community trend in advance.   
\end{abstract}





\maketitle
{\fontsize{8pt}{8pt} \selectfont
\textbf{ACM Reference Format:} \\
Jiahao Yuan, Zhao Li, Pengcheng Zou, Xuan Gao, Jinwei Pan, Wendi Ji, and Xiaoling Wang. 2022. Community Trend Prediction on Heterogeneous Graph in E-commerce. In \textit{Proceedings of the Fifteenth ACM International Conference on Web Search and Data Mining (WSDM ’22), February 21--25, 2022, Tempe, AZ, USA.} ACM, New York, NY, USA, 9 pages. https://doi.org/10.1145/3\\488560.3498522}

\section{Introduction}
With the prosperous development of e-commerce, online shopping platforms, such as Taobao and Amazon, have become an indispensable part of the modern business system. 
For customers, e-commerce platforms provide them with a rich catalog of products, help them find products of preference, and give them the great convenience of comparison with prices. 
For sellers, e-commerce platforms can assist in the product promotion and help for the insight into the market requirement to make better plans in product design, production, warehousing and sales.

In e-commerce, sellers need to produce and prepare more differentiated and trending products with the changing of users' preferences. 
For the platform, different communities have different preferences because of the user segmentation. 
It is necessary to capture the market trend in a prophetic vision, find the product elements with growth potential, and help merchants with targeted supply. 
It is important to meet the needs of consumers while empowering the platform sellers.
Therefore, it is a crucial task to capture the fashion trends in e-commerce, which can assist many downstream tasks, such as product design~\cite{yu2019personalized} and supply chain optimization~\cite{perea2003model}.
Meanwhile, users that the merchant faces are a slice of all consumers, and one item is also aimed at a specific group of users. Thus, finding the fashion trends on a specific user group is particularly significant, which we call \textbf{community trend prediction} in this paper. 
These community trends play an important role in many applications such as community search~\cite{li2015influential, wang2021efficient}, promotion recommendation~\cite{zhao2015commerce, goldenberg2020free, ji2021large}, group recommendation~\cite{cao2018attentive, wang2020group}.

\begin{figure}[!t]
    \centering
    \includegraphics[width=\columnwidth]{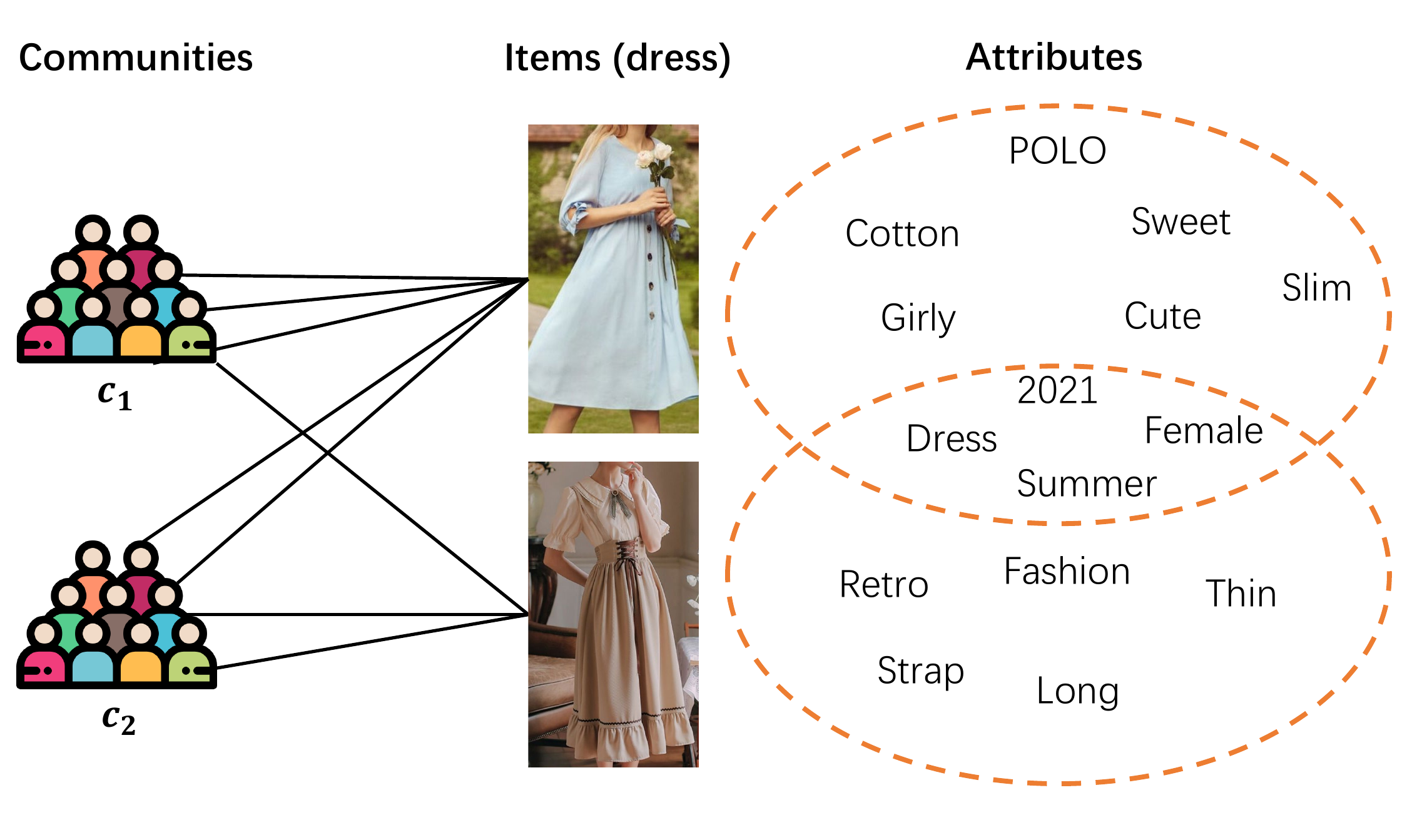}
    \caption{An example of attribute tags in different communities.}
    \label{fig:introduction}
\end{figure}
The task of community trend prediction can be based on different granularity, such as the item or the stock-keeping unit (SKU). From the perspective of users' decision-making, merchants prefer to know about the basis of users' decision-making, extract the core attributes from the products sold, and gain insight into the decision-making attributes of different communities and market segments.
For example, as depicted in Figure~\ref{fig:introduction}, there are obvious differences in the attributes of dresses purchased by different communities, such as women of different ages. 
These attribute tags can better reflect why users purchase and which the element should be paid attention to in the design.

However, the task of community trend prediction on item attributes for e-commerce brings three fundamental challenges.
First, attribute tags have almost no additional features other than the sales volume. In e-commerce platforms, a specific item usually has many related features, such as color, category, picture, description, comment, which are the basis for the online recommendation. 
As for attribute tags, they are usually derived from the features of items. Apart from its own semantic information, we do not have any other direct features to characterize it.  
Second, the semantic similarity of attribute tags in trends is difficult to capture by simple methods. Specifically, popular attributes are a set of attribute tags related to each other in the community. The attribute tags such as puff sleeves, short sleeves, sleeveless, express the different design styles on the dress sleeve. In real applications, this correlation in the community is implicit and difficult to capture.
Third, the community trend changes dynamically over time and is disturbed by external factors. For example, seasonal attributes have a huge impact on the attribute tags of clothing, and promotions have an impact on sales of all items. The periodic external factors have a more significant impact on the long-term fashion trend.

Traditionally, the staffs of fashion forecasting companies collect information of consumers' ways of living, thinking, and behaving~\cite{kim2021fashion}. Recently, some works provided an alternative data-driven way of addressing the fashion trend forecasting task~\cite{al2017fashion, ma2020knowledge}. 
In this paper, we focus on the fashion trend of the item attribute and propose a unified framework that combines the \textbf{Dy}namic evolution of \textbf{T}wo \textbf{Graph} patterns (i.e., bipartite graph and hypergraph) to predict the fashion attribute tags in the specific community (named \textbf{DyTGraph} for brevity). To the best of our knowledge, we are the first to investigate the community trend at the granularity of attribute tags in e-commerce. 

In order to overcome the challenges above, we first design a community-attribute bipartite graph on a specific item category based on the historical purchased records. 
In this bipartite graph, we use \textit{GraphSage}~\cite{hamilton2017inductive} to learn a representation of the attribute tag and capture the collaboration of different communities.
Next, another key step in our framework is to transform the bipartite graph into a hypergraph.
The hypergraph generalizes the notion of an edge in simple graphs to a hyperedge that can connect more than two nodes~\cite{bretto2013hypergraph}, which has been widely used in e-commerce ~\cite{ji2020dual, xue2021multiplex, xia2021self}.
Hence, we naturally build a hyperedge on the set of attribute tags that a community has interacted with and then introduce the hypergraph neural network to explore the correlation of these attributes.
Finally, to infer the dynamic evolution of fashion trends in the community, we utilize a recurrent neural network (RNN) to incorporate the attribute tag embedding with the corresponding sales volume at each time step to model the long- and short-term temporal patterns. 

The main contributions are summarized as follows:
\begin{itemize}
    \item In this paper, we focus on the community trend prediction in the e-commerce platform and define a novel task that mining and predicting the fashion trend at the granularity of attribute tags. We formulate this problem as a node classification task of dynamic graphs. 
    \item We propose a unified framework to address the fundamental challenges in the task. Specifically, we design a bipartite graph between communities and attribute tags to learn the representation based on historical interaction. We also introduce a hypergraph to capture the correlation among the attribute tags interacted by the specific community. 
    \item We introduce a RNN-based model that incorporates the representation of the attribute tags with sales volume to tackle the dynamic evolution in the community trend prediction.
    \item We conduct extensive experiments on three datasets from the large e-commerce platform Taobao with different communities to demonstrate the effectiveness of the proposed framework.
    
\end{itemize}

\section{Preliminaries}
In this section, we introduce the preliminary  knowledge about hypergraph and formulate the problem to address in this paper. 
\subsection{Hypergraph}
Hypergraph theory is a generalization of graph theory. The basic idea consists in considering sets as generalized edges and then in calling hypergraph the family of these edges (hyperedges)~\cite{bretto2013hypergraph}. A hypergraph is usually defined as $\mathcal{G} = \{\mathcal{V}, \mathcal{E}\}$, where $\mathcal{V}$ represents the vertex set, and $\mathcal{E}$ denotes the hyperedge set. Each hyperedge involves two or more nodes and is assigned a positive weight $W_{e}$. 
All the weights formulate a diagonal matrix $W \in \mathbb{R}^{|\mathcal{E}| \times |\mathcal{E}|}$.
The hypergraph $\mathcal{G}$ can be denoted by a incidence matrix $\mathbf{H}^{|\mathcal{V}| \times |\mathcal{E}|}$, with each entry $h(v,e)$ indicating whether a vertex $v$ is connected by a hyperedge $e$:
\begin{equation}
h(v, e) = \left \{ \begin{aligned}
     1 & \quad if \quad v \in e \\
     0 & \quad if \quad v \notin e
\end{aligned}
\right.
\end{equation}

Next, let $\mathbf{D}_e$ and $\mathbf{D}_v$ denote the diagonal matrices of the edge degrees and the vertex degrees, respectively.
For a vertex $v \in \mathcal{V}$, its degree is defined as $d(v) = \sum_{e\in \mathcal{E}}W_eh(v, e)$.
For an edge $e \in \mathcal{E}$, its degree is defined as $\delta (e) = \sum_{v\in \mathcal{V}} h(v,e)$.

\subsection{Problem Statement}
Let $C = \{c_1, c_2,..., c_N\}$ and $A = \{a_1, a_2,..., a_M\}$ denote the set of $N$ communities and $M$ attribute tags, respectively. 
The task of this paper can be regarded as a node classification, since we construct a bipartite graph and a hypergraph to capture the attribute information.
More formally, given a series of fully observed interactions $X = \{X_1, X_2,...,X_{L-1}\}$, where $X_t$ denotes all interactions between the communities and attribute tags at the month $t$ and $L$ is the length of observed time windows. 
Since we focus on the long-term popular trend, the data collection and statistics are at the granularity of the month. 
Each tuple $(c_i, a_i, s_i) \in X_t$ denotes the users belonging to the community $c_i$ have purchased the item that has the attribute $a_i$ with $s_i$ times in $t$. That is, $s_i$ denotes the sales volume. 
Noting that each item may have more than one attribute tag, so the sales on all attribute tags are not equivalent to the sales on all items.

The goal of this paper is to predict whether an attribute tag $a_i \in A$ is ranked at top-$K\%$ on the community $c_i$ in the month $L$, and did not appear in the rank list last year. In other words, the label of an attribute tag is:
\begin{equation}
y_{c_i, a_i}^{L} = \left \{ \begin{aligned}
     1, &  \quad a_i \in l^L_{c_i} \quad and \quad a_i \notin l^{L-12}_{c_i} \\
     0, & \quad others
\end{aligned}
\right.
\end{equation}
where $l_{c_i}^t$ denote the top-$K\%$ rank list on the community $c_i$ in the month $t$, which is based on the sales volume $s_i$. 
We remove the same attribute tags in last year because the trend of attribute tags is not equivalent to having the most sales. 
According to the observation of the real-world dataset, the top attribute tags are usually stable, and the classical popular elements can be directly obtained by a statistic. 
However, it is usually more desirable to find the potential attribute tags with rapid sales volume growth that cannot be easily found in advance. In this paper, we set $K=50$ without specifying.

\section{METHODOLOGY}
In this section, we present the proposed model in detail. We first summarize the  pipeline  of  the proposed framework. Then we describe the two key components.
One is to encode the attribute tag with a dynamic community-attribute bipartite graph, and the other is to exploit hypergraph to capture the correlation of the attribute tags. 
Lastly, we introduce a RNN-based model to learn the evolution of the attribute tag trend.

\begin{figure*}[t]
    \centering
    \includegraphics[width= \textwidth]{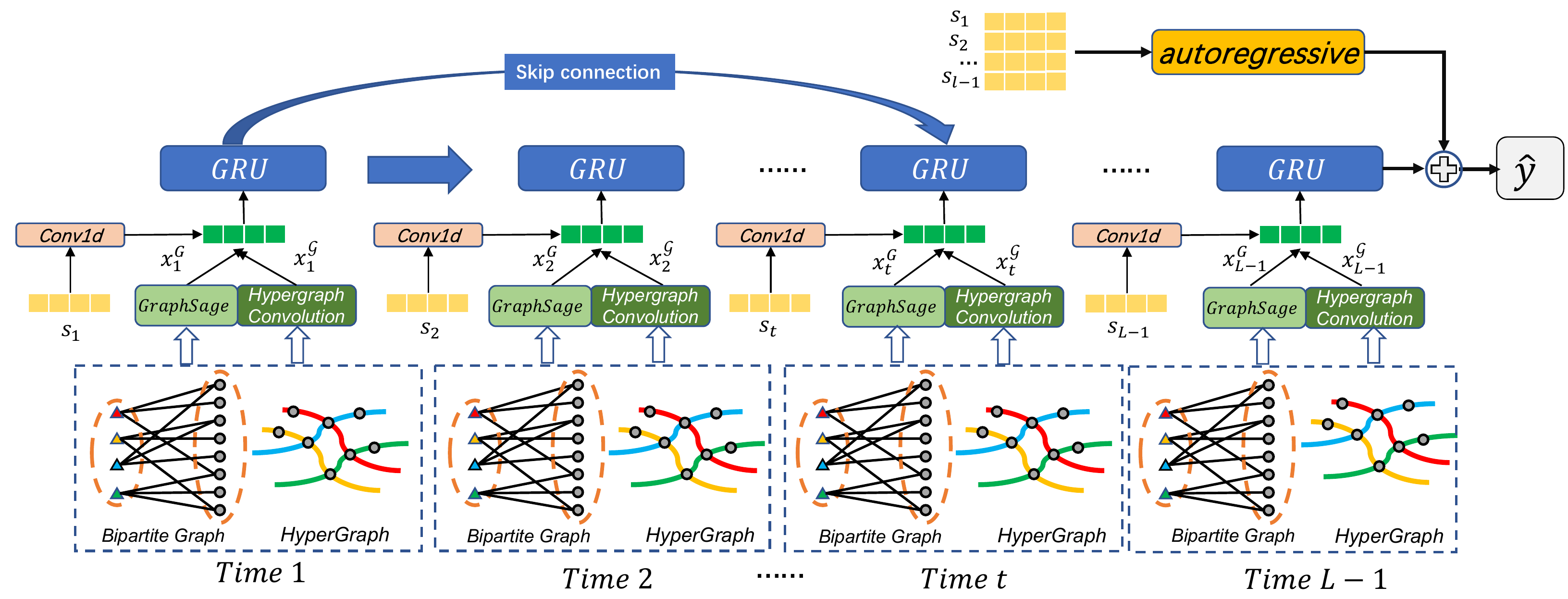}
    \caption{Overview of the proposed framework.}
    \label{Model}
\end{figure*}

\subsection{Model Overview}
The pipeline of the proposed framework is illustrated in Figure~\ref{Model}. Firstly, we transform the interaction data between communities and attribute tags in each month into a bipartite graph and a hypergraph. 
For the bipartite graph, we use $GraphSage$ to extract useful collaborative information to learn a representation embedding for each tag. Meanwhile, for the hypergraph, we utilize hypergraph convolutional neural network to capture the correlation among attribute tags and learn another embedding of it. 
Finally, we combine the two representations, and leverage the strength of a long- and short-term recurrent network to discover the evolution of the attribute trend.

\subsection{Dynamic Bipartite Graph Component}
The premise of mining fashion trend is to obtain the dynamic representation of each attribute tag at each moment. In the real-world application, it is difficult to find other features to describe the attribute tags except their own semantics. Thus, we construct a community-attribute bipartite graph based on the historical interaction and get the embedding by graph neural networks (GNNs).  

\subsubsection{Bipartite Graph Construction}
We first transform observed interactions $X=\{X_1, X_2,...,X_{L-1}\}$ into a dynamic bipartite graph $G = \{G_1, G_2,...,G_{L-1}\} $, which is an ordered sequence including $L-1$ static graph snapshots. Each graph $G_t = (C_t, A_t, E_t)$ is a weighted bipartite graph with two shared disjoint node sets $(C, A)$, and a link set $V$ at the time step $t$. Each link denotes that at least one user in the community $c_i \in C$ has purchased the item that has the attribute $a_i \in A$. The weight of each link is the sales volume $s_i$.

For example, we can divide all users into $N$ communities based on some characteristics (e.g., age, location, and gender). 
For a user in the community $c_1$, if he bought an item with attributes $(a_1, a_2, a_3)$ (e.g., sleeveless, silk and sexy dress) at the month $t$, there are three edges linking $c_1$ to the attribute tags $a_1, a_2, a_3$ in graph $G_t$, respectively. 
There is only one edge between two nodes, and the weight on the edges represents how many times the community has interacted with the attribute tag in the month. Therefore, we regard the community trend prediction as node classification rather than link prediction on the heterogeneous graph~\cite{aggarwal2014framework, li2020type}.

\subsubsection{Learning Attribute Embedding}
Before passing the graph into the proposed model, we construct two embedding matrices $M^C \in \mathbb{R}^{|C| \times d}$ for communities and $M^A \in \mathbb{R}^{|A| \times d}$ for attribute tags, where $d$ is the latent dimension.
For the node $c_k \in C_t$ and $a_j \in A_t $ in the bipartite graph $G_t$, we get their embeddings  $e_{c_k}, e_{a_j} \in \mathbb{R}^d$ by the lookup of matrix $M^C$ and $M^A$, respectively. So the initial hidden layer $h_0$ is as follows:
\begin{equation}
    h_0 = \{e_{c_k}, e_{a_j}| c_k \in C_t, a_j \in A_t\}.
\end{equation}

Next, we generate the novel attribute tag embedding with \textit{GraphSage}~\cite{hamilton2017inductive}. For the node $a_j \in A_t$, a propagation step passes information from the community node at the step $i$:
\begin{equation}
\begin{aligned}
e_{\mathcal{N}(a_j)}^i =& f_{agg}(\{e_{c_k}, \forall c_k \in \mathcal{N}(a_j)\}) \\
\hat{e}_{a_j}^{i+1} =& f_{non} (W_G \cdot concat(e_{a_j}^i, e_{\mathcal{N}(a_j)}^i)) \\
e_{a_j}^{i+1} =& norm(\hat{e}_{a_j}^{i+1})
\end{aligned}
\end{equation}
where $\mathcal{N}(\cdot)$ denotes the neighbor nodes, $W_G \in \mathbb{R}^{2d \times d}$ is the weight matrix, $f_{non}(\cdot)$ is the nonlinear activation function and $norm$ denotes the $L_2$ normalization operation. $f_{agg}$ denotes the aggregator function, which is used to aggregate information from neighbor nodes. Here, we use the inductive variant of GCN~\cite{wang2019deep}, which is:
\begin{equation}
     f_{agg}(\{e_{c_k}, \forall c_k \in \mathcal{N}(a_j)\}) = \sum_{c_k\in \mathcal{N}(a_j)} \frac{1}{d(a_j)}e_{c_k} W_{agg}
\end{equation}
where $w_{agg} \in \mathbb{R}^{d \times d}$ is the weight metrix and $d(a_j)$ denotes the degree of the node $a_j$.

The mini-batch setting can be applied in the embedding generation process. In each step, the attribute node $a_j \in A_t$ first aggregates representations of the neighbor community nodes, $\{e_{c_k}, \forall c_k \in \mathcal{N}(a_j)\}$, into a single vector  $e_{\mathcal{N}(a_j)}^i$. Then we update the embedding of $a_j$ with the vector $e_{\mathcal{N}(a_j)}^i$ to exploit the collaborative information about the community. Note that we are not taking the edge weight (i.e., the sales volume) into account here because we use it explicitly in the temporal evolution component. Moreover, since the community is determined in advance while attribute tags evolve dynamically in this paper, we take the representation of the community to be static, and do not update it with neighbor information.

Lastly, the output of this component is the last hidden state as follow:
\begin{equation}
    h_G = \{e_{c_k}, x^{G}_j| c_k \in C_t, a_j \in A_t\}
\end{equation}
where  $x^{G}_j$ denote the attribute embedding which has been updated with neighbor information in the bipartite graph.

\subsection{Dynamic HyperGraph Component}
By the bipartite graph network, the attribute nodes have aggregated information of neighbor communities to generate their representations. 
However, the bipartite graph lacks the direct connection between attribute tags and cannot model the correlation between the fashions in trend. 
Thus, we introduce an approach to transform the bipartite graph into the hypergraph to capture the correlation among attribute tags in the specific community by neural networks. 

\subsubsection{Hypergraph Construction}
For each bipartite graph $G_t = (C_t, A_t, E_t) \in G$, we transform it into a hypergraph $\mathcal{G}_t = (A_t, \mathcal{E}_t)$ where $A_t \in A$ is the node set and $\mathcal{E}_t$ denotes hyperedge set. The hypergraph $\mathcal{G}_t$ share the same attribute nodes with the bipartite graph $G_t$. From $G_t$, we regard all neighbor attribute nodes of a community node as a set, and take a hyperedge to connect them in the hypergraph $\mathcal{G}_t$. In other words, for a node $c_k \in C_t$ in $G_t$, a hyperedge is introduced to connect all nodes in $\mathcal{N}(c_k)$, i.e., the attribute nodes that are directly connected to $c_k$ by a edge in $G_t$. So the number of hyperedges in each $\mathcal{G}_t$ is equal to the number of communities $|C_t|$.

For instance, in the community-attribute bipartite network, the community $c_1$ has purchased items including the three attributes $(a_1, a_2, a_3)$, which corresponds to a hyperedge that connects these three attributes in the hypergraph. Similarly, for each bipartite graph $G_t$, we construct a hypergraph $\mathcal{G}_t$ in the time step $t$. Therefore, we get a dynamic hypergraph $\mathcal{G} = \{\mathcal{G}_1,\mathcal{G}_2,...,\mathcal{G}_{L-1}\}$.

\subsubsection{HyperGraph Convolutional Networks}
After the hypergraph construction, We apply the hypergraph convolution on each hypergraph to learn another representation of the attribute node. The hypergraph convolutional operators have borrowed ideas from the spectral theory on simple graphs to propagate the information of nodes. Firstly, the initial embedding of each node is the same as $G_t$:
\begin{equation}
    x_j^0 = e_{a_j}
\end{equation}
where we use $x_j^0$ to denote the initial embedding of the node $a_j$ in the hypergraph since there is only one type of node (attribute).

Inspired by the spectral hypergraph convolution proposed in~\cite{feng2019hypergraph}, we define our hypergraph convolution as:
\begin{equation}
    X_A^{i+1} = f_{non}(HWH^T\cdot X_A^iP^i)
\end{equation}
where $H \in \mathbb{R}^{|A_t| \times |C_t|}$ denotes the incidence matrix, $X_A$ are embeddings of all attribute nodes in the hypergraph, $P$ is the learnable weight matrix and $W \in \mathbb{R}^{|C_t| \times |C_t| }$ is a diagonal matrix for the hyperedge weight. For each hyperedge, we assign the same weight 1. Then the hypergraph convolutional operator with symmetric normalization is:
\begin{equation}
   X_A^{i+1} = f_{non}(D^{-1/2}HWB^{-1}H^TD^{-1/2}\cdot X_A^iP^i)
\end{equation}
where $f_{non}$ denotes the nonlinear activate function (i.e., ReLU function here), $D$ is the node degree matrix and $B$ is the diagonal matrix of the hyperedge degree. In the hypergraph convolution, the propagation of information has two stages. First, the multiplication operation $H^T \cdot X^i_A P^i$ aggregates all nodes on a hyperedge to get the intermediate representation of the hyperedge. Then by premultiplying $H$, each node aggregates the information of all hyperedges that it is located. 
In the proposed framework, this \textit{node-hyperedge-node} information propagation ensures that we can capture the correlations among attributes that are purchased by the specific community.

After passing the attribute embedding $X_A$ through the hypergraph convoluional layer, we get another representation of each node in the timestamp $t$, which is:
\begin{equation}
   h_{\mathcal{G}} =  \{x_j^{\mathcal{G}}|a_j \in A_t\}
\end{equation}
where $x_j^{\mathcal{G}}$ denotes the output embedding of the attribute $a_j$. 

\subsection{Dynamic Evolution Component}
Fashion trends in the community are temporal dynamics. We need to model the dynamic evolution of the attribute trend based on the two graph patterns in each time step. In e-commerce platforms, community preferences are easily affected by external factors such as season. Therefore, referring to the multivariate time series forecasting approach proposed in~\cite{lai2018modeling}, we introduce a RNN-based model to capture dynamic evolution of the attribute.   

\subsubsection{Attribute Embedding}
For each attribute, the sales volume is an essential feature of its dynamic evolution at each time step. Inspired by~\cite{zhou2021informer}, we adopt the 1-D convolutional filters (kernel width$=3$, stride$=1$) to project the sales volume into a $d$-dim vector:
\begin{equation}
    h^s_{j} = RELU (W_s \ast s_j + b_s)
\end{equation}
where $\ast$ denotes the convolution operation, $W_s$ and $b_s$ are parameters of the convolutional networks. $s_j \in \mathbb{R}^{|C_t|}$ denotes the sales for the attribute $a_j$ and each element $s_{k,j}$ in $s_j$ is the sales volume in the community $c_k$. Then the attribute embedding at each time $t$ is:
\begin{equation}
    x_j^t = (1-\alpha)x_j^G + \alpha x_j^\mathcal{G} + h_s^j
\end{equation}
where $\alpha$ is a hyper-parameter to control the fusion of information from two graph patterns.
\subsubsection{Learning Attribute Embedding Evolution}
To capture the trend evolution, we first feed the attribute embedding $x_j^t$ into a Gated Recurrent Unit (GRU)~\cite{chung2014empirical}, the hidden state of recurrent units at time $t$ is computed as:
\begin{equation}
    \begin{aligned}
    r_t &= \sigma(x_j^tW_{xr} + h_{t-1}W_{hr} + b_r) \\
    z_t &= \sigma(x_j^tW_{xu} + h_{t-1}W_{hu} + b_u) \\
    n_t &= tanh(x_j^tW_{xc} + r_t \odot (h_{t-1}W_{hc} + b_c) \\
    h_t &= (1-z_t) \odot n_t + z_t \odot h_{t-1} 
    \end{aligned}
\end{equation}
where $\odot$ is the element-wise product. The output $h_t$ is the hidden state at the time step $t$, which includes the dynamic information.

Since GRU unit usually fails to capture very long-term
correlation and ignores the periodic pattern of the sequence, we adopt another GRU structure with temporal skip-connections, which is formulated as:
\begin{equation}
\begin{aligned}
    r_t &= \sigma(x_j^tW_{xr} + h_{t-p}W_{hr} + b_r) \\
    z_t &= \sigma(x_j^tW_{xu} + h_{t-p}W_{hu} + b_u) \\
    n_t &= tanh(x_j^tW_{xc} + r_t \odot (h_{t-p}W_{hc} + b_c) \\
    h_t &= (1-z_t) \odot n_t + z_t \odot h_{t-p}
    \end{aligned}
\end{equation}
where the input $x_j^t$ is the same as the vanilla GRU unit, $p$ is the number of hidden cells skipped through and is determined for the real periodic pattern. In this way, the novel generated embedding can extend the temporal span of the information flow to capture the evolution of trends. 

Next, we combine the outputs of the vanilla GRU and skip-GRU:
\begin{equation}
    h_t^D = W^R h_t^R + \sum_{i=1}^{p-1}W_i^Sh_{t-i}^S + b  
\end{equation}
where $W^R, W_i^S \in \mathbb{R}^{d\times d}$ denote the learnable weights, $b \in \mathbb{R}^d$ is the bias vector, $h_t^R$ denotes the output of the vanilla GRU, and $h_{t-i}^S$ denotes the hidden state of the time step $t-i$. Note that we use the last period with $p$ hidden states.

Considering the important influence of the sales volume on the attribute trend in the specific community, we adopt an autoregressive (AR) ~\cite{bollerslev1986generalized} as the linear part, which is formulated as:
\begin{equation}
    \hat{s}_L^{k,j} = \sum_{t=0}^{L-1} w_t^{k,j} s_t^{k,j} + b
\end{equation}
where $w_t^{k,j}, b \in \mathbb{R}$ denote the coefficient and bias, $\hat{s}_L^{k,j}$ denotes the forecasting result of the sales volume for the attribute $a_j$ on the community $c_k$, and $s_t^{d, j}$ denotes its sales volume at the time step $t$.

\subsection{Prediction and Optimization}
\subsubsection{Prediction}
Taking the static community embedding $e_{c_k}$, the attribute embedding $h_t^D$ after dynamic evolution and the sales volume forecast $\hat{s}_L^{k,j}$ as inputs, we can estimate the group preference $\hat{r}_{k,j}$ as follows:
\begin{equation}
    \hat{r}_{k,j} = \sigma(h_t^De_{c_k}^\top + \hat{s}_L^{k,j})
\end{equation}

\subsubsection{Loss Function} After obtaining the estimated community preference on the attribute, we need to optimize the model via minimizing the loss function. In this paper, we use binary cross-entropy as the optimization objective to learn the parameters:
\begin{equation}
    \mathcal{L} = -\sum_{c_k \in C, a_j \in A} (y_{k,j}\log \hat{r}_{k,j} + (1-y_{k,j})\log (1-\hat{r}_{k,j}))
\end{equation}

\section{Experiment}
\subsection{Datasets}

The dataset used in the experiment is collected from a large e-commerce platform Taobao. We collect transaction snapshots of three top categories (Dress, Pants, and T-shirts ) on the platform from all users in Shanghai over 25 months (2019.6.17-2021.7.17). For each transaction snapshot, we extract attribute tags from the title of the purchased item to construct the attribute set. For the sake of improving retrieval results by search engines, merchants usually write a title including all important attributes of the item. Specifically, we tokenize the original long title, use named entity recognition (NER)~\cite{nadeau2007survey} in e-commerce to split the title into attribute tags. We filter out meaningless or clearly incorrect tags, only leave the tags that describe the attributes of the item itself, including style, material, fashion, color, and so on.

Next, we divide users by age and gender, as these two demographics are widely used in user research and promotion. For \textbf{\textit{Dress}} category, we investigate on female consumers, divide the primary online shopping users into seven communities based on the age (i.e., 12-18, 19-22, 23-25, 26-30, 31-35, 36-40, 41-50) and filter out attribute tags that have been purchased less than 100 times in the latest month. For \textbf{\textit{Pants}} category, the only difference is that we investigate on male consumers. For \textbf{\textit{T-shirts}} category, the difference is that we divide 14 communities according to the gender and the age (two genders with seven ages). The statistics of these three datasets are shown in Table~\ref{dataset}.

\begin{table}
\caption{Statistics of the datasets used.}
\label{dataset} 
\begin{tabular}{l|rrr}
	\toprule
	Datesets & Dress  & Pants & T-shirts\\
	\midrule
	\# users &3,895,619& 3,143,319 &5,515,647\\
	\# communities      &     7  &     7 & 14\\
	\# attributes &    3,537  &  3,917   &1,770\\
	\bottomrule
\end{tabular}
\end{table}


\subsection{Baselines and Metrics}
To the best of our knowledge, no existing algorithm can directly deal with the task of the community trend prediction. We conduct experiments of our proposed method with several popular baselines to demonstrate the effectiveness as below.
\begin{itemize}
    \item \textbf{MOM} is a statistical method of \textit{month on month}, which is a simple method widely used by the operation manager in real-world applications. In this method, the top-$K\%$ attribute tags in the current month are directly regarded as the result of the next month.  
    \item \textbf{GRU}~\cite{chung2014empirical} uses a variant of recurrent neural network. We only take the sales volume of the attribute in each community as input in this baseline for the task. 
    \item \textbf{LSTNet}~\cite{lai2018modeling} is a popular multivariate time series forecasting deep learning model. We consider the sales volume of the single attribute on different communities as the multivariate time series and feed them into the method for prediction.    
    \item \textbf{GraphSage-BI}~\cite{hamilton2017inductive, li2020hierarchical} is a method that we adopt a bipartite GraphSage proposed in~\cite{li2020hierarchical} to learn the attribute embedding from the community-attribute bipartite graph and also capture the dynamic evolution with LSTNet. 
    \item \textbf{DyTGraph-BI} is our proposed model but only uses GraphSage to extract the information from the bipartite graph.  
    \item \textbf{DyTGraph-HP} removes the bipartite graph of our proposed model, and only uses the hypergraph convolutional operator to learn the attribute embedding from the hypergraph. 
    \item \textbf{DyTGraph-GRU} only uses GRU to capture the dynamic evolution of the attribute representation. 
\end{itemize}

We adopt the area under the ROC curve (AUC) to evaluate the performance of all methods. AUC is the most popular evaluation metric on prediction tasks in research and industrial areas~\cite{krichene2020sampled}, which can be formulated as: 
\begin{equation}
    AUC = \frac{1}{|l^+||l^-|} \sum_{r\in l^+} \sum_{r' \in l^-} \mathbb{I} (r > r') 
\end{equation}
where $l^+$, $l^-$ denote the set of positive and negative samples, respectively. $\mathbb{I}(\cdot)$ is the indicator function.
AUC measures the likelihood that a random positive sample is ranked higher than a random negative sample. A larger value means a better performance.

\subsection{Experimental Setup}
In the experiment, we set the observed period to 12 for all datasets. That is to say, we use the data from the past year to predict the community trend next month.
we take the result of the last month (2021.6.17-2021.7.17) as the test set, and the penultimate month (2021.5.17-2021.6.17) as the validation set. Therefore, for each attribute tag, we have generated 11 training samples ($\{t_1, t_2,...,t_{12};t_{13}\},\\ \{t_2,t_3,...,t_{13};t_{14}\},...,\{t_{11}, t_{12},...,t_{22};t_{23}\}$).

The hyper-parameters for all methods in comparison are tuned on the validation set via gird search. We use Adam as the model optimizer, and tune the learning rate in $[0.001,0.003,0.005,0.008,0.01]$ and $\alpha$ in $[0, 0.25, 0.5, 0.75,1]$. 
Furthermore, mini-batch sizes are $64$, $16$, $64$ for Dress, Pants, and T-shirts, respectively. The embedding size $d$ is $64$, and the largest number of epochs is $100$, The skip-length $p$ of recurrent-skip layer is set as 3 for fair comparison in all datasets. We implement our model on data science workshop (DSW) \footnote{\url{https://www.aliyun.com/activity/bigdata/pai/dsw}} of Platform of Artificial Intelligence (PAI) in Alibaba Cloud.

\begin{table*}[tb]
\small
    \centering
    \caption{Performances (\%) of all methods on Dress and Pants. The highest scores are boldfaced for the proposed framework; the highest scores of baselines are underlined.}
    \label{tab:overall}
    \begin{tabular}{l|cccccccccccccccc}
    \toprule
    \multirow{2}*{Method} &  \multicolumn{7}{c}{Dress} & \multicolumn{7}{c}{Pants} \\
    \cmidrule(lr){2-9} \cmidrule(lr){10-17} 
    & 12-18 & 19-22& 23-25 & 26-30 & 31-35 & 36-40 & 41-50 & avg & 12-18 & 19-22 & 23-25 & 26-30 & 31-35 & 36-40 & 41-50 & avg \\
    \midrule  
    MOM &64.94&75.73&73.14&74.55&72.52&70.08&76.37&72.48&67.35&80.26&80.00&78.76&70.75&77.14&69.37&74.80\\
    GRU &76.54&\underline{91.22}&\underline{91.12}&86.74&84.17&\underline{87.54}&\underline{87.78}&\underline{86.44}&\underline{80.45}&89.59&93.19&88.73&86.83&84.28&80.43& 87.18\\
    LSTNet &\underline{77.63}&90.06&90.57&\underline{87.04}&\underline{84.60}&86.79&87.35&86.29&80.08&\underline{92.70}&\underline{94.47}&\underline{90.91}&\underline{89.79}&88.51&\underline{88.47}&\underline{89.28}\\
    GraphSage-BI &71.70&87.58&88.21&85.33&81.21&80.16&78.04&81.75&78.67&90.25&91.48&87.73&88.27&\underline{89.43}&85.37&87.31\\
    \midrule
    DyTGraph-BI &76.94&\textbf{91.94}&\textbf{91.37}&88.46&88.40&88.89&90.41&88.06&83.97&90.79&91.52&89.55&89.13&89.12&89.57 &89.09\\
    DyTGraph-HP &\textbf{78.68}&91.58&90.35&\textbf{89.07}&88.70&89.67&\textbf{90.83}&\textbf{88.41}&84.28&92.77&\textbf{93.58}&90.37&88.98&88.90&89.07&89.71 \\
    DyTGraph-GRU &71.78&90.41&88.87&88.04&86.91&88.06&89.55&86.23&77.88&91.60&90.29&87.68&89.32&88.43 &88.20&87.63\\
    DyTGraph &77.41&91.53&91.05&89.02&\textbf{88.85}&\textbf{89.91}&90.67&88.35&\textbf{87.32}&\textbf{92.89}&92.59&\textbf{92.50}&\textbf{90.56}&\textbf{92.46}&\textbf{90.87} &\textbf{91.31} \\
	\bottomrule
    \end{tabular}
\end{table*}

\begin{table*}[tb]
\small
    \centering
    \caption{Performances (\%) of all methods on T-shirts. The highest scores are boldfaced for the proposed framework; the highest scores of baselines are underlined.}
    \label{tab:overall2}
    \begin{tabular}{l|ccccccccccccccc}
    \toprule
    \multirow{2}*{Method}  & \multicolumn{7}{c}{T-shirts (Male)} & \multicolumn{7}{c}{T-shirts (Female)} \\
    \cmidrule(lr){2-8} \cmidrule(lr){9-15} 
    & 12-18 & 19-22& 23-25 & 26-30 & 31-35 & 36-40 & 41-50 &  12-18 & 19-22 & 23-25 & 26-30 & 31-35 & 36-40 & 41-50 & avg \\
    \midrule  
    MOM &60.98&63.32&59.98&63.21&64.32&67.36&65.54& 74.63 &64.32 &72.28&67.36&72.45&65.54&73.27 & 68.18\\
    GRU &78.16&79.41&78.05&\underline{90.04}&\underline{79.62}&\underline{91.57}&\underline{85.03}&\underline{86.98}&\underline{86.50}&\underline{85.92}&\underline{87.48}&\underline{85.81}&\underline{86.00}&\underline{84.61} & \underline{84.66}\\
    LSTNet &\underline{78.93}&\underline{80.49}&\underline{79.72}&90.02&79.20&91.08&84.72&85.58&85.91&83.85&86.02&84.88&85.75&83.69&84.27\\
    GraphSage-BI &68.49&75.92&74.04&88.61&77.46&86.88&82.97&82.08&84.39&80.49&84.34&80.24&80.90&78.55&80.38\\
    \midrule
    DyTGraph-BI &79.02&79.40&79.41&90.14&78.85&91.33&84.52&88.56&84.58&\textbf{88.40}&87.80&\textbf{89.61}&85.76&\textbf{87.93}&85.38 \\
    DyTGraph-HP &81.20&\textbf{81.91}&\textbf{81.47}&89.45&80.75&89.18&83.80&84.17&85.33&84.22&86.61&86.37&85.78&84.65&84.64 \\
    DyTGraph-GRU &77.43&77.47&80.54&88.63&\textbf{82.31}&87.68&\textbf{86.46}&86.60&85.80&87.39&87.17&87.68&79.38&85.47& 84.29 \\
    DyTGraph &\textbf{81.66}&80.98&80.87&\textbf{89.86}&81.12&\textbf{90.89}&85.30&\textbf{87.44}&\textbf{86.80}&87.43&\textbf{88.14}&88.92&\textbf{86.67}&86.16&\textbf{85.87} \\
	\bottomrule
    \end{tabular}
\end{table*}
\subsection{Experimental Results}
We compare performances over different methods on different datasets and list them in Table~\ref{tab:overall} and Table~\ref{tab:overall2}. We display results on different communities and average results of all communities.
\subsubsection{Overall Performance}
For the overall performance, we compare the proposed framework with several popular baselines, which leads to the following findings. 
 
 First, as a method which is often used in real-world applications, MOM cannot effectively discover the community trend of the item attribute. 
 The fundamental assumption of MOM is that fashion trends are inertial so that the fashion trends can be directly identified by month on month. 
 This assumption may be right in a short period but does not apply to our scene. 
 In the granularity of month, the fashion trends of dresses, pants, and T-shirts have been considerably changed, so that it is not easy to take out the popular attribute tags in the next month by MOM.
 
 Second, the sales volume is the only feature we have and is also very important for the prediction. GRU and LSTNet are two baselines based on the sequence model. We regard the sales volume of different communities as multivariate time series and predict the trend by these two models, which can effectively improve the performance. However, we observe that the proposed approaches in general outperform these two methods, which suggests that the dynamic evolution of the trend has a more complex pattern that the time series forecasting based on the sales volume cannot capture it.
 
 Third, the dynamic evolution of the community harms the performance of the trend prediction. GraphSage-BI, which also uses the bipartite graph to extract the features of attribute tags, updates the representation of the community at each time step. We find that its performance is worse than LSTNet and GRU. We infer that the representation of the community is supposed to be static since we delineate the community by some given demographic features (e.g., gender, age), which are enough to distinguish the community.   
 
 Fourth, the proposed framework, including DyTGraph and its variants, has generally achieved the best performance, which clearly demonstrates the effectiveness of utilizing two graph patterns.
 Compared with sequential models, the framework goes further to take into account the historical purchase records between communities and attribute tags, which has provided a deeper understanding of community trends. Compared with GraphSage-BI, the framework works better as it exploits the correlation of attribute tags by the hypergraph rather than just relying on the bipartite graph. 

\subsubsection{Ablation Study}
In order to measure the contribution from the different components of \textit{DyTGraph}, we compare it with three ablated versions: \textit{DyTGraph-BI}, \textit{DyTGraph-HP}, \textit{DyTGraph-GRU}.

From the results of \textit{avg} on three datasets, DyTGraph-GRU yields the worst performance, confirming that the simple sequential model such as GRU cannot fully capture the dynamic evolution of the attribute trend from two graph patterns. DyTGraph-BI has the second best performance in T-shirts, proving that the historical community-attribute bipartite graphs have indeed helped to make better predictions sometimes.
DyTGraph-HP generally is better than DyTGraph-BI, which has further demonstrated the effectiveness of explicitly modeling the correlation of attribute tags in one community, since they often share some common semantics which are hard to find.

From results on different communities, the performances of these models are slightly more complicated. In general, DyTGraph-GRU yields the worst performance, while the complete framework DyTGraph has the best performance. However, the other two models, DyTGraph-BI and DyTGraph, are not necessarily superior to each other in different communities. For instance, DyTGraph-BI achieves better performances on three communities in \textit{pants} dataset, while DyTGraph-HP is better on four communities. Therefore, the fact that the complete DyTGraph model in general performs best confirms the advantage of fusing two different graph patterns.

\subsection{Discussion of Incorporating Two Graph Patterns}
\begin{figure*}[tb]
	\centering
    \includegraphics[width= \textwidth]{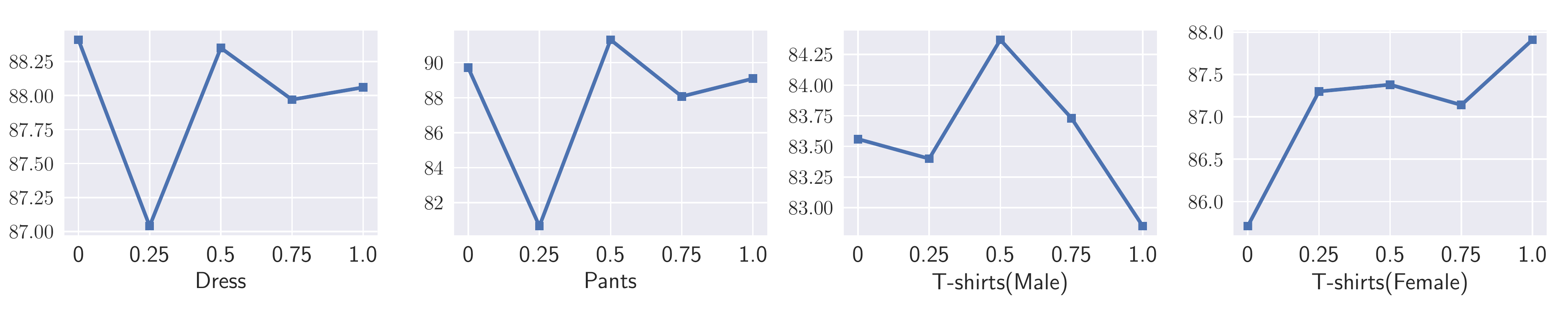}

	\caption{Performance (\%) comparison to different $\alpha$ .}
	\label{fig:alpha}
\end{figure*}

In order to investigate the effect of two graph patterns, we tune $\alpha$ in $ \{0,0.25, 0.5, 0.75, 1\}$ to control the fusion of their information. We only use average results on three datasets and report results of different gender on T-shirts dataset. 

As illustrated in Figure \ref{fig:alpha}, it is reasonable to infer that how to select the information is crucial to the final performance. Moreover, the final result is sensitive to $\alpha$. When $\alpha = 0$, we only use the hypergraph for attribute tags. When $\alpha = 1$, we only use the bipartite graph. It is worth noting that the final result is not a monotonous curve, indicating that the fusion of the two different graph patterns is complex. We set $\alpha=0.5$ for the complete model since it has the best performance in most cases.

\subsection{Case Study}

\begin{figure}[!t]
    \centering
    \includegraphics[width=\columnwidth]{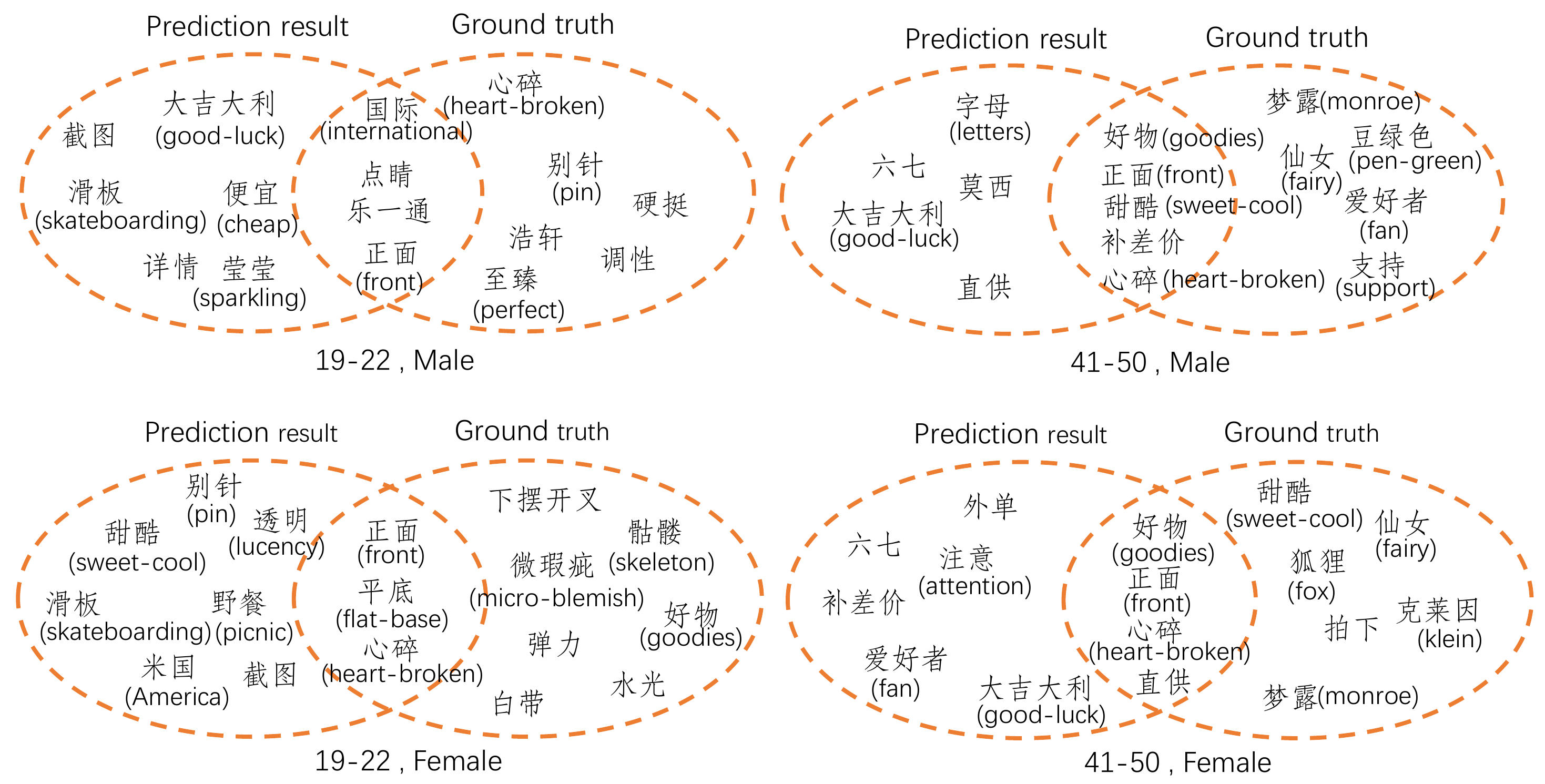}
    \caption{The case of attribute tags in trend generated by DyTGraph with the ground truth at top 10. }
    \label{fig:case}
\end{figure}

In order to intuitively demonstrate the effect of our model in the community trend prediction, we give a case study in T-shirts with  four communities. As illustrated in Figure~\ref{fig:case}, we provide the attribute tags ranked at top 10. Since the merchants provide these attribute tags themselves, the meanings of some tags cannot be translated accurately, and we try our best to translate some words. We can observe that there are some differences in attribute tags preferred by different communities. From the perspective of age, attribute tags of the two communities differ significantly. There is only one common tag "heart-broken" for males, and three common tags for females. From the perspective of gender, the younger communities have two common tags, while the relatively senior communities have six common tags. 
We infer that accounts of males in "41-50" are more likely to buy for their companions since we observe tags like "Monroe" and "fairy". The proposed model can capture the differences between different communities and find a certain number of attribute tags in trend. These tags help operation managers design different strategies for different communities.
\section{Related Works}

The research works related to us are mainly in the field of fashion trend analysis about style learning. The key issue in this field is thus how to analyze discriminative features for different styles and also learn what style makes a trend ~\cite{cheng2021fashion}. Pioneering works~\cite{59} presented an algorithm based on the stylistic coherent and unique characteristics to automatically discover visual style elements representing fashion trends for a certain season of New York. Following ~\cite{59}, many works investigate on fashion trends from a visual perspective. The paper~\cite{17} collected images from the New York Fashion Shows and New York street chic, and utilize a learning-based approach to discover fashion attributes. QuadNet~\cite{43} uses a CNN-based image embedding network to analyze fashion trends of street photos. The works ~\cite{12, 133,al2020paris} all engaged in studying the differences of fashion trends in different regions.

Further, the work~\cite{18} devised a machine learning based method considering the images of best-selling products to discover fine-grained clothing attributes in e-commerce. For the dynamic evolution of trends, the work~\cite{53} estimated users’ fashion-aware personalized ranking functions to model the visual evolution of fashion trends in recommender systems. FANCY~\cite{jeon2021fancy} developes a deep learning model that detects attributes in a given fashion image and reflects fashion professionals’ insight.

However, all of these works are to analyze fashion trends by images of items. They use the task of image classification, object detection, or popularity prediction to learn a representation and then try to extract specific fashion elements by methods such as clustering. They cannot directly predict whether a particular attribute tag will become popular in the future. Recently, KNER~\cite{ma2020knowledge} leverages internal and external knowledge in the fashion domain that affects the time-series patterns of fashion element trends. This work demonstrated the effectiveness of trends prediction by the time-series framework but still constructed a regression task for popularity prediction and required external knowledge, which cannot be directly applied in real-world scenarios. 

In this paper, we focus on the community trend prediction on item attribute without any picture information, leverage historical purchase records to establish the heterogeneous graph and hypergraph for attribute tag representations, which is more suitable for e-commerce applications.




\section{Conclusion}
In this paper, we propose a novel framework DyTGraph for community trend prediction, which considers not only the bipartite graph but also the hypergraph of attribute tags for the dynamic evolution. 
Specifically, we design a community-attribute graph at each time step to learn the collaboration of different communities and construct a hypergraph of attribute tags to exploit their associations. 
Our experiments demonstrate that the proposed model outperforms all popular baselines and find some popular tags in advance.
For future work, it would be interesting to investigate how to dynamically divide the communities according to preferences.

\begin{acks}
This work was supported by NSFC grants (No.  61972155), the Science and Technology Commission of Shanghai Municipality (20DZ1100300), Shanghai Knowledge Service Platform Project (No. ZF1213), and Shanghai Trusted Industry Internet Software Collaborative Innovation Center.
\end{acks}

\bibliographystyle{ACM-Reference-Format}
\balance
\bibliography{acmart}

\end{document}